\documentclass[prb,twocolumn,showpacs]{revtex4}
\usepackage{graphicx}
\usepackage{amsfonts}
\usepackage{amsmath}
\usepackage{amssymb}
\usepackage{bm}
\usepackage{hyperref}
\usepackage{slashed}
\begin{document}
\title{Antiferromagnetic magnons from fractionalized excitations}

\author{Rui Wang$^{1,2}$}
\author{Baigeng Wang$^{1,2}$}
\email{bgwang@nju.edu.cn}
\author{Tigran Sedrakyan$^{3}$}
\email{tsedrakyan@physics.umass.edu}
\affiliation{$^1$National Laboratory of Solid State Microstructures and Collaborative Innovation Center of Advanced Microstructures, Nanjing University, Nanjing 210093, China
}
\affiliation{$^2$Department of Physics, Nanjing University, Nanjing 210093, China
}
\affiliation{$^3$Department of Physics, University of Massachusetts Amherst, Amherst, Massachusetts 01003, USA}

\date{\today }

\begin{abstract}
We develop an approach to describe antiferromagnetic magnons on a bipartite lattice supporting the N\'{e}el state using fractionalized degrees of freedom typically inherent to quantum spin liquids.
In particular we consider a long-range magnetically ordered state of interacting two-dimensional quantum spin$-1/2$ models using the Chern-Simons (CS) fermion representation of interacting spins.
The interaction leads to Cooper instability and pairing of CS fermions, and to
CS superconductivity which spontaneously violates the continuous $\mathrm{U}(1)$ symmetry generating a linearly-dispersing gapless Nambu-Goldstone mode due to phase fluctuations.  We evaluate this mode
and show that it is in high-precision agreement with magnons of the corresponding  N\'{e}el antiferromagnet irrespective to the lattice symmetry.
Using the fermion formulation of a system with competing interactions, we show that the frustration gives raise to nontrivial long-range four, six, and higher-leg interaction vertices mediated by the CS gauge field, which are responsible for restoring of the continuous symmetry at sufficiently strong frustration. We identify these new interaction vertices and discuss their implications to unconventional phase transitions. We also apply the proposed theory to a model of anyons that can be tuned continuously from fermions to bosons.



\end{abstract}

\maketitle

\section{introduction}
Physics of two-dimensional (2D) quantum magnets has, for a long time, been in the focus of  both theorists and experimentalists \cite{Arovas,Chakravarty,Manousakis,Read,palee,Sachdev,Assa,Starykh,LeBlanc}.
Vast variety of interesting phases has been established in such systems, which emerge according to the nature of many-body correlations, symmetries, and breaking thereof.
Theoretically, physical systems exhibiting long-range magnetically ordered state are often treated using Schwinger representation of quantum spins \cite{Arovas,Assa,Ghioldi,Lmessio,weiwang}.
Ordering of spins is governed by breaking of the rotational symmetry and, upon deciding on a good choice of representation,  guided by {\em condensation} of bosonic degrees of freedom.

This framework however encounters challenges when the ground state of a magnet is a QSL. Here elementary excitations are fractionalized and no long-range order, bose-condensation, and continuous symmetry breaking take place \cite{Kalmeyer, KW1,KW2,xgwen,xgwenb,alicea,lbalents,syan,potter,erezberg,Sedrakyanb,norman,savary,kanoda,berg}.
QSLs typically support emergence of the gauge fields and absence of local order parameters.

Because the continuous symmetries are conserved in QSLs, they lead to unconventional quantum phase transitions that are beyond conventional description based on breaking of symmetries \cite{Xu,sslee,we,kuklov,kun,Ashvin,sachdev,sudbo,FvO,chalker,verstraete,CX,FB,decon}.
Theoretical understanding of such phase transitions appears to be an open and complex problem due to non-locality, topological order, and emergence of gauge theories.
The common approach to unconventional criticality is to start with the condensate and study the emergence of the gauge fields in the vicinity of the phase transition. The downside of this approach is that in the QSL state bosonic fields, condensation of which elegantly describes magnetic ordering, are fractionalized and no longer represent the fields that determine the low-energy physics.

While searching for a quantum field theory that describes the transition from a spin-liquid state with fractionalized excitations to an ordered state with Goldstone boson (e. g. magnon) excitations,
one will need to identify a unique field that should be used in describing these two fundamentally different states.

In this work,  rewinding the standard approach backwards, we start from the picture of the fractional excitations inherent to the QSL and propose that this representation can be used to describe the magnetically ordered state. Integrating out the gauge degrees of freedom, a low-energy effective action in terms of matter fields (i. e. fractionalized excitations) can be obtained.
When the magnetic order sets in, the fundamental question we ask here is the following: { how to describe massless Goldstone boson modes of the magnetically ordered state using the effective action in terms of fractionalized excitations}?

To bring the discussion to the quantitative level, let us consider the example of a transition from a chiral spin-liquid (CSL) to the planar XY N\'{e}el ordered state (although the results of the present work do not depend on this particular choice of the transition and concern the ordered state, model examples where this type of transition may happen have been reported in e.g. Refs. \onlinecite{KW1,KW2,Sedrakyanb,Sedrakyana,Sedrakyanc,we,Barkeshli,CFW93}). Amazingly, traces of emergent gauge fields and the features inherent to the two-dimensional Kitaev spin liquid and the CSL have been traced even in ladder models \cite{lHur,Petrescu}, where the precursors of these topologically ordered states start to develop. The theory of the CSL can be cast in the form of fermions coupled to U(1) CS gauge field \cite{Kalmeyer,Barkeshli,Sedrakyanb,shapo,CFW93}. On the other hand, an approach using these CS fermions \cite{Lopez,Halperin,Shankar,Jackson,Jain,Sedrakyana} to describe magnetically ordered
states was initiated in Ref. \onlinecite{Sedrakyanc}. There, instead of assuming a specific mean-field for the CS gauge field, the ordered state is obtained
non-perturbatively as the nonlocal interaction between the fermions leads to instability against Cooper pairing and to the CS superconductivity \cite{footnote}. 

Specifically, here we develop a description of magnons on an arbitrary bipartite lattice supporting the N\'{e}el state using fractionalized degrees of freedom.
The description proceeds as follows.
The CS superconductor spontaneously violates a $\mathrm{U}(1)$ symmetry as a result of the pining of the phase of wave function, and hence generates a gapless Goldstone mode due to phase fluctuation.  By carefully evaluating the two Goldstone modes, i.e., the magnons in N\'{e}el AFM order and the phase fluctuation in CS superconductor, we found that the two modes have dispersions that are in good agreement
with each other, both qualitatively and quantitatively.

In order to examine the dependence on lattice symmetry, two typical bipartite lattice, i.e., hexagonal and square lattices are considered.
Even though the pairing mechanism is found to be different in the hexagonal and square lattice, they display exactly the same excitation spectrum in the region when the CS superconductor is stabilized.
This verifies universality of the description in the sense of its lattice symmetry independence.
Moreover, we apply the proposed mechanism to a generalized model of anyons with tunable statistics from fermionic to bosonic\cite{CFW93} excitations. The mean-field theory of CS superconductor requires a critical $e_c$, indicating that the system could exhibit a phase transition from a gapless Dirac liquid for $e<e_c$ to an ordered state with bosonic excitations for $e>e_c$.


Furthermore, the situation changes dramatically when a frustrating spin-spin interaction is present, and a transition into CSL
in planar XY antiferromagnets becomes advantageous\cite{Sedrakyanb,Sedrakyana,Sedrakyanc,we,Sheng15}.
It appears that in this situation, the  system in its fermionized representation forms nonlocal six-leg interaction vertices,
which become dominating in the regime of strong frustration.  With the occurrence of unconventional phase transition in the system,
it would be interesting to examine the
nature of instabilities driven by these interaction vertices, as they must restore the continuous $U(1)$ symmetry.
One advantage offered by our approach is that the fermion representation can be used both to describe the ordered phase and the
spin-liquid, thus has a potential to offering an additional route to understanding the unconventional criticality.

The rest of this paper is organized as follows.
In the next section we derive the magnon spectra corresponding to $s=1/2$ XY antiferromagnets on bipartite lattices including a square and honeycomb ones. In Sec.III we consider the CS superconductor corresponding to these lattices and compute the dispersion relations of Nambu-Goldstone modes.
In Sec.IV we compare the spectra of these modes with those of magnons and show that  they are in a good agreement with each other no matter what the lattice symmetry is. In Sec.V we discuss the implications of the above theory of breaking of the continuous $U(1)$ symmetry
to unconventional phase transitions separating ordered CS superconducting phases and quantum spin-liquids. In particular we identify the operators in the Hamiltonian that are responsible for restoring of the continuous symmetry. We present conclusions in Sec.VI while the technical details of computations are presented in Appendices A-C.



\section{Magnons in planar N\'{e}el antiferromagnets}
Consider a spin-1/2 XY antiferromagnet (AFM) with nearest-neighbor interaction on a bipartite lattice. In general, the Hamiltonian reads as,
\begin{equation}\label{eq1}
  H=J\sum_{\langle ij\rangle}\hat{S}^x_i\hat{S}^x_j+\hat{S}^y_i\hat{S}^y_j,
\end{equation}
where the sum over nearest-neighbor (NN) indices $i,j$ is implied.
In the following we firstly consider the example of a square lattice, but the results can be straightforwardly generalized to any bipartite lattice with no frustration. The ground state of the model Eq.(\ref{eq1}) has been verified to be the in-plane AFM N\'{e}el  order \cite{kennedy,dhlee}, being subject to quantum fluctuations. The purpose of our theory below is not to study the ground state itself, but to investigate the low-energy excitation.  The N\'{e}el order breaks the in-plane $\mathrm{U}(1)$ spin rotation symmetry, which leads to low-energy gapless Goldstone mode, namely, spin wave or magnons.
On a square lattice, two sublattices, $A$ and $B$, with opposite in-plane spin orientations emerge in the ground state. With an in-plane rotation of coordinate, the spin orientation can always be set along $\hat{x}$, hence, $\hat{S}^x_{\alpha}|\alpha\rangle=s^x_{\alpha}|\alpha\rangle$, where $|\alpha\rangle$ denotes the spin eigenstate at $\alpha=A,B$ sublattice, with $s^x_{A}=\frac{1}{2}$, $s^x_{B}=-\frac{1}{2}$.  To increase or decrease $s_x$, one can introduce the spin raising or descending operators, $ S^{\pm}=S^z\mp iS^y$. This setup motivates one to introduce a modified Holstein-Primakov transformation by defining spin operators at each lattice site as $S^+_{Ai}=(\sqrt{1-a^{\dagger}_ia_i})a_i,
S^{-}_{Ai}=a^{\dagger}_{i}\sqrt{1-a^{\dagger}_ia_i}$, and $S^{+}_{Bi}=b^{\dagger}_i\sqrt{1-b^{\dagger}_ib_i}, S^-_{Bi}=(\sqrt{1-b^{\dagger}_ib_i})b_i$, with $a_i$, $b_i$ being the annihilation operators for magnons at site $i$. Assuming that magnon fluctuations are much smaller than the total spin, the Hamiltonian describing magnon excitations then can be cast as
\begin{equation}\label{eq2}
\begin{split}
  H&=\sum_i{E_ib^{\dagger}_ib_i+E_ia^{\dagger}_ia_i}-t\sum_{i,j}(a^{\dagger}_ib_{i+\boldsymbol{\delta}_j}+a_ib^{\dagger}_{i+\boldsymbol{\delta}_j})\\
  &+\Delta\sum_{i,j}(a^{\dagger}_{i}b^{\dagger}_{i+\boldsymbol{\delta}_j}+b_{i+\boldsymbol{\delta}_j}a_i)-\frac{1}{4}JNZ,
\end{split}
\end{equation}
where $N$ is the number of sites of each of the sublattices, $\boldsymbol{\delta}_j$ denote the lattice vectors corresponding to NN bonds, and $Z=4$ is the number of nearest-neighbor sites on a square lattice. Eq.\eqref{eq2} includes two flavors of magnons with the same onsite energy, $E_i=\frac{ZJ}{2}$, the NN hopping as well as the NN pairing between different magnons with the amplitude $t=\Delta=\frac{J}{4}$. After performing Bogoliubov transformation, the energy spectrum of magnons can be obtained. The two lower bands have the dispersion, $E_{\mathbf{k}}=\pm\frac{J}{2}\sqrt{Z^2-Z|\phi_{\mathbf{k}}|}$, where $\phi_{\mathbf{k}}=\sum_{j}e^{i\mathbf{k}\cdot\boldsymbol{\delta}_j}$. The dispersion is gapless and doubly degenerate at $\Gamma$ point. Expansion in the vicinity of $\Gamma$ yields the linear dispersion of the spin wave or magnons, $E_{\mathbf{k}}=Jak=\frac{1}{2}v_Fk$, where $a$ is the lattice constant shown in Fig.1(a). Here we have introduced the ``Fermi velocity" $v_F=2Ja$. The obtained doubly degenerate gapless magnon spectrum is in accordance with the hardcore boson description of the N\'{e}el ordered state, where bosons condense at $\Gamma$ point in the momentum space. By comparison to the Heisenberg model \cite{fransson}, the dispersion is decreased by a factor $\sqrt{2}$. This is because the kinetic energy of a spin wave is restricted due to the absence of $s_z$ component in the XY model.


\section{Nambu-Goldstone mode in a Chern-Simons superconductor}
To put forward a different description of magnons using fractionalized excitations, in the following, we derive the Nambu-Goldstone modes of CS superconductors and compare them with corresponding magnons.
In the following subsection, taking the square lattice as an example, we give a detailed demonstration of how a CS superconductor is formed.

\subsection{CS superconductor on a square lattice}




Here we start with an effective description of a long-ranged-ordered state of $s=1/2$ XY antiferromagnet  on a square lattice.
Firstly, we note that the spin raising/lowering operators, $S^{\pm}$, in Eq.(\ref{eq1}) can be represented using  fermionic degrees of freedom attached to a flux (sometimes referred to as {\em string}) operator \cite{Jackson,Lopez,Shankar,Halperin,Jain,Sedrakyana,Sedrakyanb}, $U^{\pm}_i$, corresponding to the same cite:
$  S^{\pm}_i=f^{\pm}_iU^{\pm}_i$,
where the string operator,
$  U^{+}_i=e^{ie\sum_{i^{\prime}\neq i}\mathrm{arg}(\mathbf{R}_i-\mathbf{R}_{i^{\prime}})\hat{n}_{i^{\prime}}}$.
Here $e$ is the CS charge that represents the number of attached fluxes, and the fermion occupation number is related to spin operators as $\hat{n}_{i}=f^{\dagger}_{i}f_{i}=\hat{S}^z+1/2$. Note that to conserve the correct commutation relation of spin operators, $e$ has to be an odd-integer: $e=2l+1$,
$l \in \mathbb{Z}$.
This is a key property that makes the fermion representation of spin-1/2 rising/lowering operators exact.

In the interest of generalization of the
proposed theory, and for further possible practical applications, we shall consider more general "anyon creation/annihilation" operators, $ \bar{S}^{\pm}=f^{\pm}_iU^{\pm}_i$,  whose
CS charge in flux operators, $ U^{+}_i$, is allowed to be any real number, $e\in \mathbb{R}$. This will allow us to treat the charge as a continuous parameter in all further considerations, while when $e$ changes from zero to one, the statistics of anyons changes from Fermi to Bose.
To select the sectors of the Hilbert space of $s=1/2$ operators under consideration in Eq.(\ref{eq1}), that correspond to replacement $ \bar{S}^{\pm} \rightarrow  {S}^{\pm}$, one should
add an additional $\delta$-function in the functional integral representation of the partition function, ${\mathcal Z}$, of the $s=1/2$ XY antiferromagnet: ${\mathcal Z} = {\frac{1}{2\pi}} \int_{-\infty}^{\infty}de\int_{-\infty}^{\infty}dm\exp\left\{im[e-(2l+1)]\right\}{\mathcal Z}[e]$. Here
${\mathcal Z}[e]$ corresponds to the XY model with generalized anyon operators\cite{CFW93} $\bar{S}^{\pm}$. Throughout this article we will omit the expression
giving $\delta$-function and will consider the partition function of anyon model with charge $e$, ${\mathcal Z}[e]$, noting that choosing $e=(2l+1)$ will precisely reproduce the state of  $s=1/2$ model. 

By using the above transformation, one can recast Eq.\eqref{eq1} as a theory of fermions coupled to the CS gauge field \cite{Fradkin,Sedrakyanb}.
\begin{figure}[tbp]
\includegraphics[width=3.4in]{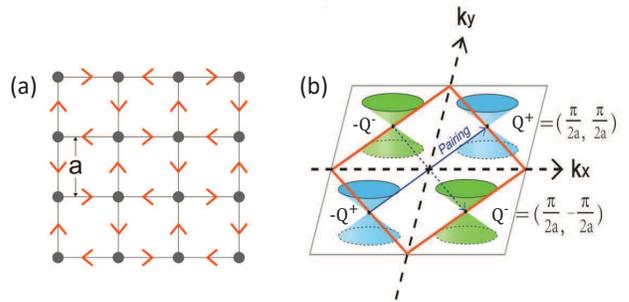}
\label{fig1}
\caption{(color online). (a) The checkerboard $\pi$-flux pattern in the square lattice. The red arrows represents a $\pi/4$ ($-\pi/4$) phase when the trajectory goes along the (reversed) direction of the arrows. (b) The low-energy spectrum of the CS fermions in square lattice. The orange square denotes the Brillouin zone. The gauge field induces a intravalley interaction (blue arrows) that favors pairing condensation at $\Gamma$ point.}
\end{figure}
The planar N\'{e}el AFM order yields two sublattices, $A$ and $B$, with opposite spin orientation. Correspondingly, in the fermionized description, there emerges a checkerboard $\pi$ flux pattern in each square plaquette \cite{Affleck}. As is shown by Fig.1(a), the nearest-neighbor plaquettes have opposite fluxes, $\pi$ and $-\pi$.
Accounting for this $\pi$ flux pattern leads to the following action of CS-fermions,
\begin{equation}\label{eq23s}
\begin{split}
S_{F}&=\int dt\sum_{\mathbf{r}} f^{\dagger}_{\mathbf{r}}(i\partial_t+A^0_{\mathbf{r}})f_{\mathbf{r}}\\
&-J\sum_{\mathbf{r},j}[f^{\dagger}_{\mathbf{r}}e^{i\mathbf{A}_{\mathbf{r}}\cdot\boldsymbol{\delta}_j}e^{i\phi_{\delta_j}}f_{\mathbf{r}+\boldsymbol{\delta}_j}+H.C.],
\end{split}
\end{equation}
where $\phi_{\delta_j}=\pm\frac{\pi}{4}$, with the sign $\pm$ being determined by the direction of arrows in a plaquette in Fig.1(a). $\mathbf{A}_{\mathbf{r}}$ is the gauge field arising from string operators in the CS-fermion representation, and it satisfies $\mathbf{A}_{\mathbf{r}}\cdot\boldsymbol{\delta}_j=e\sum_{\mathbf{r}^{\prime}}[\mathrm{arg}(\mathbf{r}-\mathbf{r^{\prime}})-\mathrm{arg}(\mathbf{r}+\boldsymbol{\delta}_j-\mathbf{r^{\prime}})]\hat{n}_{\mathbf{r}^{\prime}}$. $A^0_{\mathbf{r}}$ is the Lagrangian multiplier to ensure the Stocks theorem  of the gauge field \cite{Sedrakyanb}. Now one can firstly turn off the gauge field, and make Fourier transformation to the momentum space. This leads to
\begin{equation}\label{eq24s}
\begin{split}
  S_{F}&=\int dt\sum_{\mathbf{k},\alpha}f^{\dagger}_{\mathbf{k}\alpha}i\partial_tf_{\mathbf{k}\alpha}-2J\sum_{\mathbf{k}}[f^{\dagger}_{\mathbf{k}A}(e^{i\pi/4}\cos k_xa\\
  &+e^{-i\pi/4}\cos k_ya)f_{\mathbf{k}B}+H.C.],
\end{split}
\end{equation}
where we have used $f_{\mathbf{k}\alpha}$ and $\alpha=A,B$ to denote the spinless CS fermions on A and B sublattice. From Eq.\eqref{eq24s}, we know that in the absence of a net magnetization along $z$-axis, CS fermionization yields a half filled fermionic system subjects to the checkerboard  $\pi$ flux. The corresponding Hamiltonian gives a gapless linear spectrum near $Q^{\pm}=(\pi/2a,\pm\pi/2a)$. Since we are interested in low-energy excitations, one can expand the energy near the Dirac points and introduce a momentum cutoff $\Lambda$. The low-energy effective Hamiltonian reads as $H_0=\sum_{\lambda=\pm}H^{(\pm)}_0$, where

\begin{equation}\label{eq25s}
  H^{(\pm)}_0=-v_F\sum_{\mathbf{k}}f^{(\pm)\dagger}_{\mathbf{k}A}e^{i\pi/4}(k_x\mp ik_y)f^{(\pm)}_{\mathbf{k}B}+H.C.,
\end{equation}
where we use $f^{(\lambda)}_{\mathbf{k}\alpha}$ with $\alpha=A,B$ and $\lambda=\pm$ to denote the annihilation operator of CS fermions at  sublattice $\alpha$ and valley $\lambda$.
The additional global phase $e^{i\pi/4}$ can be removed by a gauge transformation. In the basis of $f^{(\pm)}_{\mathbf{k}}=[f^{(\pm)}_{\mathbf{k}A},f^{(\pm)}_{\mathbf{k}B}]^T$, which represents the fermion spinor at the valley $Q^{\pm}$, the Hamiltonian is then cast into:
\begin{equation}\label{eq26s}
  H_0=v_F\sum_{\mathbf{k}}f^{(+)\dagger}_{\mathbf{k}}\boldsymbol{\sigma}\cdot\mathbf{k}f^{(+)}_{\mathbf{k}}+v_F\sum_{\mathbf{k}}f^{(-)\dagger}_{\mathbf{k}}\boldsymbol{\sigma}^{T}\cdot\mathbf{k}f^{(-)}_{\mathbf{k}},
\end{equation}
From Eq.\eqref{eq26s}, it is known that in the absence of the gauge field, the phase is a ``semimetal" fermionic state with two inequivalent gapless Dirac points $Q^{\pm}$ (Fig.1(b)).
In real space, upon introducing the minimal coupling of the gauge field, the action of the XY model
based on the Chern-Simons fermion representation can be decomposed into $S_F=S_0+S_{int}$, where $S_0=S^{(+)}_0+S^{(-)}_0$, with
\begin{equation}\label{eq27s}
\begin{split}
  S^{(\pm)}_0&=\int d\mathbf{r}dt f^{(\pm)\dagger}_{\mathbf{r}}[i\partial_t-v_F\boldsymbol{\sigma}^{(T)}\cdot(-i\boldsymbol{\nabla})]f^{(\pm)}_{\mathbf{r}}
\end{split}
\end{equation}
%
and $S_{int}=S^{(+)}_{int}+S^{(-)}_{int}$, with
\begin{equation}\label{eq28s}
  S^{(\pm)}_{int}=e\int d\mathbf{r}dt f^{(\pm)\dagger}_{\mathbf{r}}\sigma^{(T)\mu}A_{\mu,\mathbf{r}}f^{(\pm)}_{\mathbf{r}}
\end{equation}
%
where we have introduced $\sigma^{\mu}=(\sigma^0,-v_F\sigma^1,-v_F\sigma^2)$. Here and in what follows, we introduce the convention that $\boldsymbol{\sigma}^{(T)}=\boldsymbol{\sigma}$ for $Q^{+}$ valley, while $\boldsymbol{\sigma}^{(T)}=\boldsymbol{\sigma}^{T}$ for $Q^{-}$ valley. Recalling that the gauge field
dynamics of the gauge field is governed by a Chern-Simons term, $S_{CS}$\cite{Sedrakyanb}, our next step is to integrate out the gauge field taking into account $S_{CS}$. It is convenient to do so in the momentum space, where the Chern-Simons action of the gauge field reads as, $S_{CS}=e\int \frac{d\omega d^2q}{(2\pi)^3}A_{\mu,\mathbf{q}}D^{-1\mu\rho}_{\mathbf{q}}A_{\rho,-\mathbf{q}}$, where
$D^{-1\mu\rho}_{\mathbf{q}}=\alpha q^{\mu}q^{\rho}+\frac{i}{4\pi}\epsilon^{\mu\nu\rho}q_{\nu}$ is the inverse of the gauge-field propagator.
After fixing the gauge by taking $\alpha\rightarrow\infty$, one can integrate out the gauge field in $S_{int}$ and $S_{CS}$, generating effective interaction between CS fermions. Before doing that, we note that since $Q^{\pm}=(\pi/2a,\pm\pi/2a)$, if one considers the intervalley interaction, then the condensation will not take place at the $\Gamma$ point, i.e., the resulting Cooper pairs favored by the interaction (see below) will exhibit nonzero total momentum
$(\pi/a,0)$. This finite momentum pairing state has been considered in other semimetal systems \cite{Bednik,Zhou}, where it was shown to be energetically ruled out by the zero-momentum Bardeen-Cooper-Schrieffer (BCS) state. Apart from this, we are interested in the condensation at $\Gamma$
point where the correspondence to hardcore boson condensation can be made. Because of this reason we will concentrate on the intravalley interaction rather than the intervalley one for the square lattice. This is different from that in the honeycomb lattice, where the intervalley interaction is responsible for the instability of CS fermions with respect to forming zero-momentum BCS Cooper pairs.

Upon integrating out the gauge field, one generates a non-local
fermion-fermion interaction, which is mediated by the dynamic CS flux attachment.
In the momentum representation, the induced intravalley coupling is found to be

\begin{equation}\label{eq33s}
\begin{split}
  S_{int}&=\sum_{\lambda=\pm}\int dt\int \frac{d\mathbf{k}d\mathbf{k}^{\prime}d\mathbf{q}}{(2\pi)^6}\\
  &\times[V^{(\lambda)\alpha\alpha^{\prime}\beta^{\prime}\beta}_{\mathbf{q}}f^{(\lambda)\dagger}_{\mathbf{k}\alpha}f^{(\lambda)\dagger}_{\mathbf{k^{\prime}+q}\alpha^{\prime}}f^{(\lambda)}_{\mathbf{k^{\prime}},\beta^{\prime}}f^{(\lambda)}_{\mathbf{k+q}\beta},
\end{split}
\end{equation}
where the interaction vertices for the two valleys are obtained as,
\begin{equation}\label{eq34s}
V^{(\pm)\alpha\alpha^{\prime}\beta^{\prime}\beta}_{\mathbf{q}}=-\pi i e v_F\epsilon_{ij}(\sigma^{i(T)}_{\alpha\beta}\delta_{\alpha^{\prime}\beta^{\prime}}-\delta_{\alpha\beta}\sigma^{i(T)}_{\alpha^{\prime}\beta^{\prime}})A^j_{\mathbf{q}}
\end{equation}
where $A^j_{\mathbf{q}}=q^j/|q|^2$, and the momentum is measured from the {\em respective Dirac point}. $\epsilon_{ij}$ is the antisymmetric Levi-Civita tensor. Comparing the terms for the two valleys in Eq.\eqref{eq33s}, it is obvious that the only difference for the two valleys amounts in a replacement of the Pauli matrices (in the sublattice space) by their transposed ones. Therefore, the physics near $Q^{-}$ and $Q^{+}$ is symmetric and the state is degenerate. Since the corresponding Hamiltonian does not contain any valley mixing terms, we arrive at the equivalent representation of the XY model on a square lattice (Eq.\eqref{eq1}) in the continuum limit using CS fermions. It is given by the direct sum of two contributions,  $H=H^{(+)}+H^{(-)}$,  where
\begin{equation}\label{eq36s}
\begin{split}
  H^{(\pm)}&=v_F\sum_{\mathbf{k}}f^{(\pm)\dagger}_{\mathbf{k}\alpha}\boldsymbol{\sigma}^{(T)}_{\alpha\beta}\cdot\mathbf{k}f^{(\pm)}_{\mathbf{k}\beta}\\
  &+\sum_{\mathbf{k},\mathbf{k}^{\prime},\mathbf{q}}V^{(\pm)\alpha\alpha^{\prime}\beta^{\prime}\beta}_{\mathbf{q}}f^{(\pm)\dagger}_{\mathbf{k}\alpha}f^{(\pm)\dagger}_{\mathbf{k^{\prime}+q}\alpha^{\prime}}f^{(\pm)}_{\mathbf{k^{\prime}},\beta^{\prime}}f^{(\pm)}_{\mathbf{k+q}\beta}.
\end{split}
\end{equation}
%
$H^{(\pm)}$ is the Hamiltonian for $Q^{\pm}$ valley correspondingly, where the Pauli matrices $\boldsymbol{\sigma}$ denoting the sublattice space are transposed for $Q^-$ valley. Interestingly enough, we note that while in the case of the honeycomb lattice the intervalley interaction dominates and favors condensation at $\Gamma$ point \cite{Sedrakyanc}, here in case of the square lattice the important role is played by intravalley interaction, see Fig.1(b). The intervalley interaction in square lattice gives rise to the Cooper pair condensation with finite momentum and  will be energetically ruled out \cite{Bednik,Zhou}. The key difference of the instabilities is due to different lattice symmetries, which result in different locations of the Dirac points.

Below we will investigate Cooper pairing instability around $Q^+$ point of the Brillouin zone. The result corresponding to the valley $Q^-$ will be the same. To proceed, we perform the Hubbard-Stratonovich decomposition by introducing fluctuating bosonic field in the particle-particle channel, with $ \Delta^{\alpha\alpha^{\prime}}_{\mathbf{k}}=\sum_{\mathbf{k}^{\prime}}V^{\beta\beta^{\prime}\alpha\alpha^{\prime}}_{\mathbf{k-k^{\prime}}}\langle f^{(+)}_{\mathbf{k}^{\prime}\beta}f^{(+)}_{-\mathbf{k}^{\prime}\beta^{\prime}}\rangle.$
Then, the interaction is decoupled to
\begin{equation}\label{eq38s}
\begin{split}
  H_{int}&=\sum_{\mathbf{k}}\Delta^{\alpha\alpha^{\prime}}_{\mathbf{k}}f^{(+)\dagger}_{-\mathbf{k}\alpha}f^{(+)\dagger}_{\mathbf{k}\alpha^{\prime}}+\Delta^{\alpha\alpha^{\prime}\star}_{\mathbf{k}}f^{(+)}_{\mathbf{k}\alpha^{\prime}}f^{(+)}_{-\mathbf{k}\alpha}\\
  &-\sum_{\mathbf{k},\mathbf{k}^{\prime}}\Delta^{\alpha\alpha^{\prime}}_{\mathbf{k}}V^{\alpha\alpha^{\prime}\beta^{\prime}\beta-1}_{\mathbf{k-k^{\prime}}}\Delta^{\beta\beta^{\prime}}_{\mathbf{k}^{\prime}}.
\end{split}
\end{equation}
Neglecting the fluctuation of the bosonic field, one can write down the mean-field Hamiltonian. Since CS fermions are spinless, their $s$-wave pairing is excluded. Due to the symmetry of order parameters as analysed in Appendix A, the natural mean-field ansatz is determined as, $\hat{\Delta}^{11}_{\mathbf{k}} = \Delta_{0x\mathbf{k}}-i\Delta_{0y\mathbf{k}}$ and $\hat{\Delta}^{12}_{\mathbf{k}} = \Delta_{3\mathbf{k}}. $
Furthermore, since the interaction between CS fermions mediated by the gauge field still exhibits p-wave nature, similar to the honeycomb lattice \cite{Sedrakyanc}, we have, $\Delta_{0x\mathbf{k}}=\Delta_{0k}\frac{k_x}{k}$ and $\Delta_{0y\mathbf{k}}=\Delta_{0k}\frac{k_y}{k}$. For square lattice, the order parameter in particle-particle channel has different symmetry in the sublattice degrees of freedom compared to the honeycomb lattice \cite{Sedrakyanc}. This is because of the different form of the induced interactions (Eq.\eqref{eq34s}), which is again a result of the different position of the Dirac points of the spectrum.

Finally, in the basis $\psi_{\mathbf{k}}=[f^{(+)}_{\mathbf{k}A},f^{(+)}_{\mathbf{k}B},f^{(+)\dagger}_{-\mathbf{k}A},f^{(+)\dagger}_{-\mathbf{k}B}]^T$, the mean-field Hamiltonian is cast in the form,
\begin{equation}\label{eq39s}
  H_{BdG}=v_F\boldsymbol{\sigma}\cdot\mathbf{k}\frac{\tau^3+1}{2}+v_F\boldsymbol{\sigma}^{T}\cdot\mathbf{k}\frac{1-\tau^3}{2}+\tau^+\hat{\Delta}_{\mathbf{k}}+\tau^-\hat{\Delta}^{\dagger}_{\mathbf{k}},
\end{equation}
with the $2\times2$ pairing potential in sublattice space, $\hat{\Delta}_{\mathbf{k}}=\Delta_{0x\mathbf{k}}\sigma^0-i\Delta_{0y\mathbf{k}}\sigma^3+i\Delta_{3\mathbf{k}}\sigma^2$.
This Hamiltonian constitutes the effective CS superconductor description on a square lattice. The Pauli matrices $\boldsymbol{\tau}$ denote the Nambu space and $\tau^{\pm}=(\tau^1\pm i\tau^2)/2$.
$\hat{\Delta}_{\mathbf{k}}$ suggests two order parameters defined as $\hat{\Delta}^{12}_{\mathbf{k}}=\Delta_{3k}$ and $\hat{\Delta}^{11}_{\mathbf{k}}=\Delta_{0k}\frac{k^-}{k}$. The self-consistent equations of the two order parameters $\Delta_{3k}$ and $\Delta_{0k}$ are derived in Appendix B, yielding
\begin{eqnarray} \label{eq89nna}
  \Delta_{0k} &=& \frac{ev_F}{4}\sum_{a=\pm}\int^k_0dk^{\prime}k^{\prime}\frac{\Delta_{3k^{\prime}}}{kE^a_{k^{\prime}}}, \\
  \Delta_{3k} &=& \frac{ev_F}{4}\sum_{a=\pm}\int^{\Lambda}_kdk^{\prime}\frac{av_Fk^{\prime}+\Delta_{0k^{\prime}}}{E^a_{k^{\prime}}}.
\end{eqnarray}
For small $k$ with $k<J/v_F$, the self-consistent equations can be analytically  solved upon Taylor expansion in terms of $k$.
To leading order in $k$, it is found that the nontrivial solution exists for $e>e_c$, with $e_c=4/\sqrt{3}$. The nontrivial order parameter justifies the superconductor instability of the CS fermions, as a result of the intermediate gauge field. To the second order in $k$, the self-consistent solution becomes almost exact in the whole Brillouin zone (BZ), which follows from comparison with the result of lattice computation \cite{Sedrakyanc}, and therefore one can set $\Lambda$ to be up to half of the BZ (one valley), $\pi/2a$, allowing to remove the $\Lambda$-dependence in the final results. The remaining higher order $O(k^3)$ corrections bring about error bar in the following calculation.


\subsection{Nambu-Goldstone mode}
Now we shift our attention to the effect of phase fluctuations in CS superconductivity.
We firstly introduce a local phase field coupled to the superconducting order parameter, $\hat{\Delta}_{\mathbf{k}}e^{i\theta(\mathrm{r},\tau)}$, with $\hat{\Delta}_{\mathbf{k}}$ being a $2\times2$ matrix. Even though there are two independent order parameters $\Delta_{0k}$ and $\Delta_{3\mathbf{k}}$ in the matrix, it can be shown that only one phase field is allowed in order to satisfy gauge invariance. Then, the imaginary-time functional action of CS superconductor can be cast in the following form:
\begin{equation}\label{eq11}
\begin{split}
  S&=\int d\tau\sum_{\mathbf{k}}\psi^{\dagger}_{\mathbf{k}}[\partial_{\tau}-v_F\boldsymbol{\sigma}\cdot\mathbf{k}\frac{\tau^3+1}{2}-v_F\boldsymbol{\sigma}^{T}\cdot\mathbf{k}\frac{1-\tau^3}{2}\\
  &-\hat{\Delta}_{\mathbf{k}}e^{2i\theta}\tau^+-\hat{\Delta}^{\dagger}_{\mathbf{k}}e^{-2i\theta}\tau^-]\psi_{\mathbf{k}},
\end{split}
\end{equation}
After performing the gauge transformation, $\psi_{\mathbf{k}}\rightarrow\psi^{\prime}_{\mathbf{k}}=U\psi_{\mathbf{k}}$, with $U=e^{-i\theta}(\tau^3+1)/2+e^{i\theta}(1-\tau^3)/2$, the phase field $\theta$ will be shifted to the diagonal terms in the Green's function (GF). The GF can then be decomposed into a phase-independent free part and a self-energy due to the phase field $\theta$, i.e.,
$G^{-1}_{\mathbf{k},\theta}=[G^{0}_{\mathbf{k}}]^{-1}-\Sigma_{\theta}$. The action of the phase field, which describes its dynamics, can be obtained by integrating out the fermionic spinor field $\psi_{\mathbf{k}}$ in Eq.\eqref{eq11}. Expanding the self-energy order by order and completing the functional integral yield $ S_{\theta}=-\mathrm{Trln}[G^{0}_{\mathbf{k}}]^{-1}+\mathrm{Tr}(G^{0}_{\mathbf{k}}\Sigma_{\theta})+\frac{1}{2}\mathrm{Tr}(G^{0}_{\mathbf{k}}\Sigma_{\theta}G^{0}_{\mathbf{k}}\Sigma_{\theta})$.
After accounting for each of the three terms, the final action turns out to be rotational invariant. As derived in Appendix C, we show that it has the standard form,
\begin{equation}\label{eq13}
  S_{\theta}=\int d\tau d\mathbf{r}[c_1(i\partial_{\tau}\theta)^2+c_2(\nabla\theta)^2],
\end{equation}
with $c_2/c_1=v^2_FI_2/I_1$, where $I_2$ and $I_1$ are dimensionless integrals determined from the self-consistent solution of the order parameters $\Delta_{3\mathbf{k}}$, $\Delta_{0k}$. The latter are, in turn, functions of the CS charge $e$. As the result, the dispersion of the phase fluctuation is obtained as, $E_{\mathbf{k}}=\sqrt{I_2/I_1}v_Fk$ (see Appendix C3), where the linearity coefficient $\sqrt{I_2/I_1}$ is a function of $e$ .

\begin{figure}[tbp]
\includegraphics[width=3.4in]{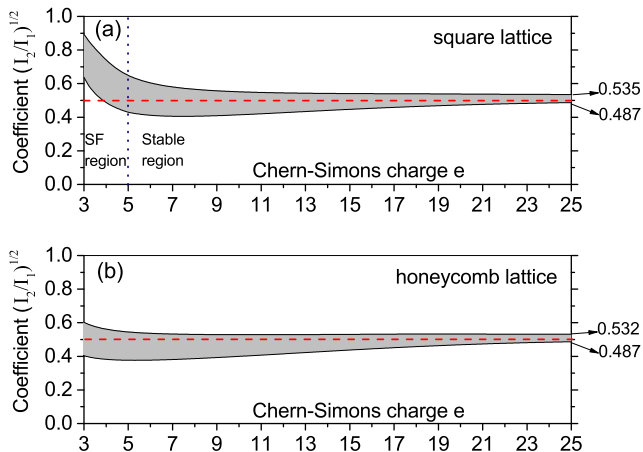}
\caption{(color online). The evaluated linearity coefficients, $\sqrt{I_2/I_1}$, of the Nambu-Goldstone modes of CS superconductors on square (a) and hexagonal (b) lattices. The red dashed line at $0.5$ denotes the linearity coefficient of the spin-wave dispersion corresponding to the N\'{e}el AFM. The grey shaded region shows the error bar due to higher order $O(k^3)$ corrections in the self-consistent solution.}
\end{figure}

\section{Comparison}

At this point we would like to remind the reader that the spin-wave dispersion obtained in the context of the N\'{e}el AFM is given by $E_{\mathbf{k}}=\frac{1}{2}v_Fk$. Qualitatively, we see that both Nambu-Goldstone mode in CS superconductivity (that possibly describes the N\'{e}el AFM state) and magnons in the N\'{e}el AFM have linear dispersions proportional to $v_F$. Quantitatively, one can further compare the linearity coefficients of the dispersions in both cases. In the magnon picture, the coefficient is a constant, $1/2$, in units of $v_F$ irrespective of the lattice symmetry (both for square and honeycomb lattices). However, in the CS superconductor picture, the coefficient $\sqrt{I_2/I_1}$ depends on the CS charge, $e$. To compare the linearity coefficients we trace the $e$-dependence of the physical low-energy excitation and compute the coefficient $\sqrt{I_2/I_1}$ as a function of $e$. The result is shown in Fig.2(a), where the shaded region shows the calculated coefficient of the Nambu-Goldstone mode with the error bar given by the width of the latter. As we see, the coefficient firstly drops with increasing $e$,   but then it shows a very weak dependence on $e$ and saturates at a finite value for larger $e$. We refer to the former and latter regions as to strongly fluctuating (SF) and the stable region respectively. Indeed, the interaction vertex
$V^{\alpha\alpha^{\prime}\beta^{\prime}\beta}_{\mathbf{q}}$ is proportional to $e$, which determines the strength of the pairing interaction.  When the CS superconductor lies in the SF region,  even though the superconductivity is stabilized and the state
has lower energy as compared to the normal state, it does not lie deep in the symmetry breaking phase so that the phase fluctuation is stronger (and thus the error bar of the computation appears to be larger). Because of this reason the linearity coefficient for smaller $e$ in SF region is rather large as compared to the one corresponding to magnons. In the SF region, it is expected that the mean-field results for the CS superconductor will deviate from the N\'{e}el AFM state, where the magnon fluctuations are assumed to be negligible compared to the total spin (see Sec.II).  When increasing $e$, the interaction becomes stronger, the CS superconductor becomes more stable and therefore the phase fluctuation is gradually suppressed. This effect is evidenced by high-precision numerics with decreased error bar and finally the Nambu-Goldstone mode acquires a saturated dispersion in the stable region. We also note that the results in Fig.2 are obtained by considering $e$ as a continuous  parameter, however, with the constraint condition from the $\delta$-function in the functional integral representation of the partition function (see Sec.III A), $e$ is automatically pinned to odd integer $e=2l+1$ to truthfully represent the original spin model.  Hence only dispersion coefficients that correspond to odd integer values of $e$ represent the physical results correspond to the XY spin model Eq.\eqref{eq1}. Except for the case $e=3$ where the strong fluctuation makes the CS superconductor less precise in mean-field level, the dispersion of phase mode remains very weakly dependent on $e$ for any $e=2l+1$. This agrees with the fact that the CS charge $e$ is a physically redundant degrees of freedom when performing feminization of the spin (hardcore boson) operators.

Numerically found dispersion of the CS superconductor is depicted in Fig.2(a) for the square lattice. It is found that the phase fluctuation of the stable CS superconductor displays the saturated dispersion $E_{\mathbf{k}}=uv_Fk$, with $u=0.511(\pm0.024)$. This result agrees remarkably well with that of the dispersion of magnons with $E_{\mathbf{k}}=0.5v_Fk$. Importantly, we have applied the above formalism to the XY model on the honeycomb lattice as well. The magnon spectrum is found to be $E_{\mathbf{k}}=0.5v^{\prime}_Fk$ as is shown by the dashed line in Fig.2(b), with $v^{\prime}_F=\sqrt{3}\epsilon J/2$  and $\epsilon$ the lattice constant of the two triangular sublattices \cite{Sedrakyanc} (Fig.3). The dispersion of the phase fluctuations is found to saturate at $E_{\mathbf{k}}=uv^{\prime}_Fk$, with $u=0.509(\pm0.023)$. Therefore, the found agreement between Nambu-Goldstone mode of the CS superconductor and the magnon mode of the N\'{e}el AFM stands equivalently well in both lattices, suggesting the lattice-independence of our theory.

The CS superconductor mean-field has its implications to the XY model Eq.~(\ref{eq1}) with
anyon operators $ \bar{S}^{\pm}$ and partition function ${\mathcal Z}[e]$ with continuously varying charge $e$. Considering ${\mathcal Z}[e]$ instead of ${\mathcal Z}$, the fermionized model from Eq.\eqref{eq1} is generalized from hardcore bosons to anyons with tunable statistics (depending on $e$). This tunable anyonic statistics can be clearly seen from the CS representation of the $ \bar{S}^{\pm}$ operators, i.e.,  $\bar{S}^{\pm}_i=f^{\pm}_iU^{\pm}_i$,
where the string operator, $  U^{+}_i=e^{ie\sum_{i^{\prime}\neq i}\mathrm{arg}(\mathbf{R}_i-\mathbf{R}_{i^{\prime}})\hat{n}_{i^{\prime}}}$ has a continuous parameter $e$ . As proved in Appendix D, the braiding between the two states, $\bar{S}^{\dagger}_i|0\rangle$ and $\bar{S}^{\dagger}_j|0\rangle$ with $|0\rangle$ being the vacuum state, gives rise to a statistical angle $\gamma=\pi+e\pi$, with $\pi$ the contribution from the braiding of the CS fermions and $e\pi$ the ``Berry phase" due to the CS flux attachment. It is clear that only for odd integer $e=2l+1$, the $\bar{S}^{\pm}$ operators return back to the hardcore bosons, while for a generic $e$ (or for $0<e<1$), the CS fermion representation produces the anyonic excitations with the statistics tuned by $e$.

While, as we see from the above discussion, the CS superconductor describes well the antiferromagnetic magnons of bosonic  $e=2l+1$ state, it implies a topological phase transition to a Dirac liquid at $e= e_c$. For anyonic values of the charge, $0<e<1$, the superconducting mean-field essentially breaks down suggesting a weakly interacting gapless Dirac liquid state. This an interesting result but the validity of this approximation needs to be studied on the matter of fluctuations and be verified numerically ({e.g.} employing diagrammatic Monte-Carlo approach). We would like to point however that it is consonant with the flux-smearing mean-field of Ref.~\onlinecite{CFW93} for the anyon model in an external magnetic field, suggesting that the
behavior of the state can be fully independent of $e$.


\section{Unconventional phase transitions}
\begin{figure}[btp]
\includegraphics[width=3.2in]{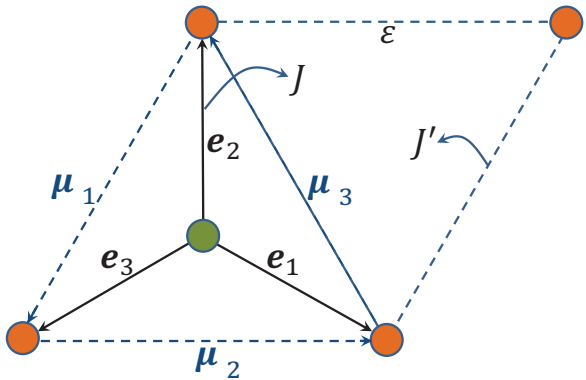}
\caption{(color online). The unit cell of the honeycomb lattice with $J$ and $J^{\prime}$ interactions. NN and NNN vectors $\mathbf{e}_j$ and ${\bm \mu}_j$ are shown while $\epsilon$ is the lattice constant of triangular sublattices.}
\end{figure}
From the above comparison of the Goldstone modes, we see that the CS superconductor is an ordered state that shares almost identical low lying excitations as the N\'{e}el AFM state.  This suggests that the mechanism of breaking of the rotational symmetry in the later can be alternatively reformulated using the fermionic language as breaking of the gauge symmetry in the CS superconductor.  With occurrence of a sufficiently strong competition and frustration in interactions however in some situations a quantum spin-liquid or a paramagnet is stabilized. These states preserve (or restore) the continuous $U(1)$ symmetry.
As we mentioned in the introduction, an interesting question is to understand the nature of these frustrating antiferromagnetic interactions using the CS fermion representation.

To this end we consider an XY model with NN, $J$, and next-NN, $J'$, spin-exchange interactions on a honeycomb lattice: $ H[J,J']=J\sum_{\langle ij\rangle}\left(\hat{S}^x_i\hat{S}^x_j+\hat{S}^y_i\hat{S}^y_j\right)+ J'\sum_{\langle\langle ij\rangle\rangle}\left(\hat{S}^x_i\hat{S}^x_j+\hat{S}^y_i\hat{S}^y_j\right)$.
We note that previous numerical \cite{zzhu,Ciolo,Carrasquilla,Varney} and analytical \cite{Sedrakyanb}  results
have shown that the rotational $U(1)$ must be preserved in the ground state of this system for sufficiently strong, but not very strong frustrations, $0.36 \gtrsim J'/J\gtrsim 0.2$.
Fermionization of  spin-half operators enables alternative formulation of the model in the form of fermions coupled to the CS string operators
\begin{equation}\label{eq14a}
\begin{split}
H[J,J']&=J\sum_{\mathbf{r},j}[f^{\dagger}_{\mathbf{r}}e^{i\mathbf{A}_{\mathbf{r}}\cdot\boldsymbol{e}_j}f_{\mathbf{r}+\boldsymbol{e}_j}+H.C.]\\
&+J'\sum_{\mathbf{r},j}[f^{\dagger}_{\mathbf{r}}e^{i\mathbf{A}_{\mathbf{r}}\cdot\boldsymbol{\mu}_j}f_{\mathbf{r}+\boldsymbol{\mu}_j}+H.C.],
\end{split}
\end{equation}
where $\boldsymbol{e}_j$ and $\boldsymbol{\mu}_j$ are the NN and next-NN  bond vectors of the honeycomb lattice (see Fig.3).
Thus one obtains a half-filled fermion system with the Fermi level consisting of two inequivalent Dirac points $K^{\pm}$ because of the
honeycomb lattice structure. Here we restrict ourselves exclusively to half-filling, which means that the number of
fermions equals to the number of sites on either of the two sublattices.
Thus, as before, one can expand in the vicinity of $K^{\pm}$ and  perform gauge tranformation\cite{Sedrakyanc}
to remove the string operators from NN and next-NN hopping terms of the Hamiltonian.
 This procedure results in a covariant derivative $\hat{\mathbf{k}}\rightarrow\hat{\mathbf{k}}+\mathbf{A}_{\mathbf{r}}$, justifying the minimum coupling used to derive Eq.\eqref{eq27s}.  
 The bare Hamiltonian reads as $H_0[J,J']=\sum_{\lambda=\pm}H_0^{(\lambda)}[J,J']$, where
 \begin{equation}\label{eqq15a}
   H_0^{(\pm)}[J,J']=\sum_{\mathbf{k}}f^{(\pm)\dagger}_{\mathbf{k}}\left(\frac{k^2}{2m}\sigma^0\pm v^{\prime}_F\mathbf{k}\cdot\boldsymbol{\sigma}^{(T)}\right)f^{(\pm)}_{\mathbf{k}}
 \end{equation}
As in Eq.~(\ref{eq26s}), we used the notation $f^{(\lambda)}_{\mathbf{k}}$ with $\lambda=\pm$ to represent the fermion spinor field at the $K^{\pm}$ valley correspondingly. Also, we remind that $\boldsymbol{\sigma}^T$ is defined as $\boldsymbol{\sigma}^T=(\sigma_x,-\sigma_y)$ and $\mathbf{k}$ is the 2D momentum vector measured from the Dirac points. An effective mass $m$ emerges in $H_0[J,J']$ due to the nonzero $J'>0$ and is given by $m=2/\left(3J^{\prime}\epsilon^2\right)$. Following the same procedure used to derive Eq.\eqref{eq33s}, we minimally couple the CS gauge field $\mathbf{A}_{\mathbf{r}}$ to $H_0[J,J']$ and then integrate out the gauge field. The intermediate gauge field is found to induce a number of different competing interaction terms in addition to the bare Hamiltonian $H_0$. The full fermionized Hamiltonian of the $J-J'$ model acquires the form $H[J,J']=H_0[J,J']+H^{(1)}_{int}+H^{(2)}_{int}+H^{(3)}_{int}$, where $H^{(1)}_{int}$ is given by
\begin{equation}\label{eq15a}
  H^{(1)}_{int}=\sum_{\mathbf{r},\mathbf{r}^{\prime},\lambda,\rho}V^{(1)\alpha\beta\alpha^{\prime}\beta^{\prime}}_{\mathbf{q}}f^{(\lambda)\dagger}_{\mathbf{k}_1\alpha}f^{(\lambda)}_{\mathbf{k}_1+\mathbf{q}\beta}
  f^{(\rho)\dagger}_{\mathbf{k}_2\alpha^{\prime}}f^{(\rho)}_{\mathbf{k}_2-\mathbf{q}\beta^{\prime}},
\end{equation}
where the interaction vertex $V^{(1)\alpha\beta\alpha^{\prime}\beta^{\prime}}_{\mathbf{q}}$ reads as
\begin{equation}\label{eq16a}
  V^{(1)\alpha\beta\alpha^{\prime}\beta^{\prime}}_{\mathbf{q}}=-i\pi ev^{\prime}_F\epsilon_{ij}(\sigma^i_{\alpha\beta}\delta_{\alpha^{\prime}\beta^{\prime}}
  +\delta_{\alpha\beta}\sigma^{iT}_{\alpha^{\prime}\beta^{\prime}})A^j_{\mathbf{q}}.
\end{equation}
The term $H^{(1)}_{int}$ is the effective interaction corresponding to the XY model on the honeycomb lattice\cite{Sedrakyanc}, unlike Eq.\eqref{eq33s} which corresponds to the square lattice symmetry. This density-current type interaction follows upon integrating out the CS gauge field  in the XY model on the honeycomb lattice and yields $p$-wave superconducting instability. The corresponding state is again a CS superconductor, which breaks the continuous gauge symmetry.


We now proceed with including an antiferromagnetic $J^{\prime}$, that leads to frustration in the XY model and competing interactions in the fermionized representation. When  parameter $J^{\prime}$ is nonzero but small, nothing essential is expected to happen with the CS superconducting state, and the Bogoliubov type mean-field treatment applied above should work well. One expects that the CS superconducting phase, initially favored by $\sim J$ term only, will be destroyed however by the frustration\cite{zzhu,Ciolo,Carrasquilla,Varney,Sedrakyanb} at $J^{\prime}/J \gtrsim 0.2$. In the CS fermion representation, the frustration is exactly manifested by the CS gauge field coupled to NNN terms.
Integrating out the gauge field one obtains two more interaction vertices for fermions, that compete with  Eq.~(\ref{eq16a}).
From the term $(\mathbf{k}+\mathbf{A}_{\mathbf{r}})^2/2m$ in Eq.\eqref{eqq15a}, the two nonlocal interaction terms $H^{(2)}_{int}$ and $H^{(3)}_{int}$ can be derived, where $H^{(2)}_{int}$ is given by
\begin{equation}\label{eq21b}
   H^{(2)}_{int}=\sum_{\mathbf{k}_1,\mathbf{k}_2,\mathbf{q}}\sum_{\lambda,\rho} \frac{\mathbf{k}_1\cdot\tilde{\mathbf{A}}_{\mathbf{q}}}{m}f^{(\lambda)\dagger}_{\mathbf{k}_1}f^{(\lambda)}_{\mathbf{k}_1+\mathbf{q}}
  f^{(\rho)\dagger}_{\mathbf{k}_2}f^{(\rho)}_{\mathbf{k}_2-\mathbf{q}},
\end{equation}
and $H^{(3)}_{int}$ is a contribution containing a six-leg interaction vertex,
\begin{equation}\label{eq22b}
\begin{split}
   H^{(3)}_{int}&=\sum_{\mathbf{q},\mathbf{q}^{\prime},\mathbf{k}_1,\mathbf{k}_2,\mathbf{k}_3}\sum_{\lambda,\rho,\delta}\frac{\tilde{\mathbf{A}}_{\mathbf{q}}\cdot\tilde{\mathbf{A}}_{\mathbf{q}^{\prime}}}{2m}\\ &\times f^{(\lambda)\dagger}_{\mathbf{k}_1}f^{(\lambda)}_{\mathbf{k}_1-\mathbf{q}-\mathbf{q}^{\prime}} f^{(\rho)\dagger}_{\mathbf{k}_2}f^{(\rho)}_{\mathbf{k}_2+\mathbf{q}}f^{(\delta)\dagger}_{\mathbf{k}_3}f^{(\delta)}_{\mathbf{k}_3+\mathbf{q}^{\prime}}.
\end{split}
\end{equation}
In Eqs.~(\ref{eq21b}) and (\ref{eq22b}) the vector potential $\tilde{\mathbf{A}}_{\mathbf{q}}$ is defined by $\tilde{A}^i_{\mathbf{q}}=2\pi ie\epsilon^{ij}q^j/|q|^2=2\pi ie\epsilon^{ij}A^j_{\mathbf{q}}$. These two additional interactions describe the effect of $J^{\prime}$ in the CS fermion language.
We believe that these new interaction terms, that to our best knowledge weren't studied before in the literature,
will lead to novel quantum phase transitions possibly involving fractional statistics of low-lying excitations.

When $J^{\prime}$ is small compared to $J$,
the terms $H^{(2)}_{int}$ and $H^{(3)}_{int}$ will not alter the stability of the  CS superconductor, which is determined by dominating $H^{(1)}_{int}$.
 With increasing $J^{\prime}$, the effect of $H^{(2)}_{int}$ and $H^{(3)}_{int}$ will become  comparable to $H^{(1)}_{int}$, and the mean-field decoupling introduced in Sec.\uppercase\expandafter{\romannumeral3} will become inapplicable. From the numerical evidence \cite{zzhu,Ciolo,Carrasquilla,Varney}, one expects possibly a continuous transition
 to a non-uniform state that preserves the $U(1)$ symmetry. It has been recently proposed to be a CSL, which also supports a charge density wave (CDW) order in terms of CS fermions\cite{Sedrakyanb}, stabilizing a finite expectation value of $\langle \hat{n}_{A}-\hat{n}_{B}\rangle$ in the region $0.36\gtrsim J^{\prime}/J\gtrsim 0.2$.

 It is natural for strong interactions to induce a CDW. It is not obvious a priori how these new interaction terms restore the $U(1)$ symmetry, but,
 since the procedure of deriving these interaction terms is formally exact, one expects this to happen at  $J^{\prime}/J\gtrsim 0.2$.
  The novelty in our present discussion is that we reformulate the model with quantum phase transition in the fermion representation and demonstrate that it should exactly capture the phases of the frustrated XY model.
     Moreover, it is interesting to observe that though the terms in Eq.~\eqref{eq21b} and  (\ref{eq22b}) are  invariant under the time reversal symmetry (TRS)   transformation (since the full $J-J^{\prime}$ XY Hamiltonian is time-reversal invariant),
it explicitly contains vector $\tilde{\mathbf{A}}_{\mathbf{q}}$ in scalar product $\mathbf{k}_{1}\cdot\tilde{\mathbf{A}}_{\mathbf{q}}$,
thus enabling stabilization of $2\pi e\epsilon^{ij}\nabla^iA^j_{\mathbf{q}}$
in real space. The latter is equivalent to the {\em smearing of the flux tubes} mean-field approximation used in the composite fermion theory of the half-filled Landau level and used in describing the CSL.  Therefore we conclude by pointing that  $H^{(2)}_{int}$ is potentially able to spontaneously break TRS
by generating the CS field  and giving rise to a CSL. This enables  $J^\prime$  to be a tuning parameter that drives the quantum phase transition between CSL and the
magnetically ordered state.

In our proposed method, this transition is interpreted as the phase transition from the CS superconductor to the CSL state driven by $H^{(2)}_{int}$ and $H^{(3)}_{int}$, as schematically depicted in Fig.~4. The process can be qualitatively understood as follows. In the CS superconductor, the low energy excitation is the phase fluctuation mode of the Cooper pairs. Besides, at finite temperatures, vortices in the SC state should also emerge which effectively correspond to the spin vortices in the XY model. With increasing $J^\prime$, the nonlocal interactions $H^{(2)}_{int}$ and $H^{(3)}_{int}$ intend to gradually break the Cooper pairs formed by CS fermions. After crossing the quantum critical point (QCP), the mean-field theory of the CS superconductor is completely violated, and therefore the CS fermions released from the Cooper pairs fill up the energy spectrum and form a Fermi surface, similar to the spinon Fermi surface in QSLs. On the other hand, the nonlocal interactions $H^{(2)}_{int}$, $H^{(3)}_{int}$ spontaneously break TRS and renormalize the CS fermions in a way similar to that in the fractional quantum Hall state, generating effective flux attachment which further gives rise to fractional excitations, i.e., the semions in the CSL.

Even though the complete study of the effect of two new vertices is yet to be performed, the present fermionic formulation of the problem sets a convenient ground for
controlled Feynman diagram calculations of correlation functions, including, {e.g.}, scalar chirality, and other Wilson-loop-based characteristics of quantum spin-liquids near unconventional criticality both analytically and using the powerful diagrammatic Monte-Carlo technique\cite{dmc1,dmc2,dmc3,dmc4}.


\begin{figure}[btp]
\includegraphics[width=3.2in]{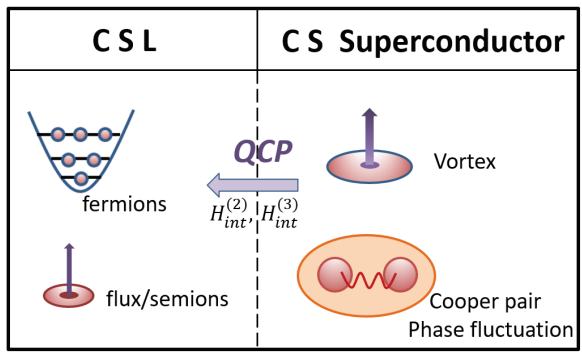}
\caption{(color online) Schematic representation of  low-energy excitations of the CS superconductor and the CSL state. The vortex in the CS superconductor effectively represents the spin vortex of the XY model. The nonlocal interactions  $H^{2}_{int}$, $H^{3}_{int}$ drive the CS superconductor through the quantum critical point into the CSL state, whose low energy excitations are semions and the "released" CS fermions from the Cooper pairs.  }
\end{figure}


\section{Conclusions}
We have developed a scheme describing the breaking of continuous symmetry within the low-energy theory of fractionalized degrees of freedom.
The approach describes antiferromagnetic magnons on a bipartite lattice using CS fermion representation of $s=1/2$ operators.
To describe magnetic ordering using CS fermions, one will have to develop a scenario for Bose-condensation of hard-core bosons.
Qualitatively, a natural candidate state is the condensation of Cooper pairs of CS fermions, leading to the CS superconductivity.
We consider a long-range magnetically ordered state of  spin$-1/2$ XY magnets that is given by
CS superconductivity which spontaneously violates the continuous $U(1)$ symmetry.
We have shown that the linearly-dispersing gapless Nambu-Goldstone mode of the CS superconductor is in  remarkably good
agreement with magnons of the XY antiferromagnet.  We have also shown that the agreement of the two Nambu-Goldstone modes is independent of lattice symmetries by considering the problem on the square and honeycomb lattices.  Another important observation is the weak $e$-dependence in the stable region of the CS superconductor state. This agrees well with the fact that the CS charge $e$ is physically redundant for any odd integer values in the CS fermionization formalism.

Theses observations suggest an interesting proposal that the N\'{e}el AFM order in the bulk of the spin-1/2 system could be described equivalently well by the CS superconductor state in the language of CS fermions. To exactly identify the equivalence between the two seemingly different phases, more evidences should be obtained following other theoretical routes, e.g., by evaluating and comparing the wave function of the two states, which can be an interesting topic for future investigation. Once this identification being proved, the new understanding of the AFM order can enjoy various applications. For example, the spin transfer problem between AFM materials  can be studied by model of CS superconductor junctions. Another possible application is the heavy fermion superconductors and high $T_c$ superconductors. Since the AFM background is a CS superconductor, the superconductivity of itinerant electrons can be understood as a result of the proximity effect of the CS superconductor.
Last, we generalize the above scheme and apply it to study frustrated magnets. Our results for the $J-J^{\prime}$ XY model on honeycomb lattice suggests an interesting perspective, i.e., the quantum phase transition from AFM to QSLs is a consequence of nonlocal interaction  between CS fermions.

We believe that the above results and the formulated approach can further stimulate a recipe and general guidelines for studies of low-lying excitations in magnetic phases, unconventional quantum phase transitions between QSL and magnetically ordered states. It would be potentially interesting to investigate the application of the proposed theory in other topics such as spin transport, heavy fermion superconductors, and high $T_c$ superconductors.

\begin{acknowledgments}
T.A.S. is grateful to A. Kamenev and V. Galitski for illuminating discussions.
This work was supported by the National Key  R\&D Program of China (Grant No. 2017YFA0303200), and by NSFC under Grants No. 11574217 and No. 60825402.
T.A.S. acknowledges startup funds from UMass Amherst.
\end{acknowledgments}

\appendix


\section{Symmetry of the superconducting order parameter}

In this appendix, we shift our attention to examining the symmetry of the interaction vertices. Explicitly writing down each term in Eq.\eqref{eq34s}, we can obtain the following equations.
\begin{equation} \label{eq38ma}
  \begin{split}
  V^{1211}_{\mathbf{q}} &= \pi ev_FA^{+}_{\mathbf{q}},    \\
  V^{1222}_{\mathbf{q}} &= \pi ev_FA^{-}_{\mathbf{q}},     \\
  V^{2111}_{\mathbf{q}} &= -\pi ev_FA^{+}_{\mathbf{q}},    \\ 
  V^{2122}_{\mathbf{q}} &= -\pi ev_FA^-_{\mathbf{q}},   \\
  V^{1112}_{\mathbf{q}} &= -\pi ev_FA^-_{\mathbf{q}},   \\
  V^{1121}_{\mathbf{q}} &= \pi ev_FA^-_{\mathbf{q}},   \\
  V^{2212}_{\mathbf{q}} &= -\pi ev_FA^{+}_{\mathbf{q}}, \\
  V^{2221}_{\mathbf{q}} &= \pi ev_FA^{+}_{\mathbf{q}}, 
  \end{split}
\end{equation}
where $A^{\pm}_{\mathbf{q}}=q_x\pm iq_y/q^2$. Besides, we can arrive at the following equations by expanding the self-consistent equation of the superconductor order parameter.
\begin{eqnarray}
  \Delta^{11}_{\mathbf{k}}=& Tr\sum_{\mathbf{k}^{\prime}}[V^{1112}_{\mathbf{k-k^{\prime}}}e^{-\beta H_{MF}}f_{\mathbf{k}^{\prime},1}f_{-\mathbf{k}^{\prime},2\nonumber}\\
  &+V^{1121}_{\mathbf{k-k^{\prime}}}e^{-\beta H_{MF}}f_{\mathbf{k}^{\prime},2}f_{-\mathbf{k}^{\prime},1}], \\ \label{eq38na}
  \Delta^{22}_{\mathbf{k}}=& Tr\sum_{\mathbf{k}^{\prime}}[V^{2212}_{\mathbf{k-k^{\prime}}}e^{-\beta H_{MF}}f_{\mathbf{k}^{\prime},1}f_{-\mathbf{k}^{\prime},2}\nonumber\\
  &+V^{2221}_{\mathbf{k-k^{\prime}}}e^{-\beta H_{MF}}f_{\mathbf{k}^{\prime},2}f_{-\mathbf{k}^{\prime},1}]. \label{eq38nb}
\end{eqnarray}
\begin{eqnarray}
  \Delta^{12}_{\mathbf{k}} =& Tr\sum_{\mathbf{k}^{\prime}}[V^{1211}_{\mathbf{k-k^{\prime}}}e^{-\beta H_{MF}}f_{\mathbf{k}^{\prime},1}f_{-\mathbf{k}^{\prime},1}\nonumber\\
  &+V^{1222}_{\mathbf{k-k^{\prime}}}e^{-\beta H_{MF}}f_{\mathbf{k}^{\prime},2}f_{-\mathbf{k}^{\prime},2}], \\ \label{eq38nna}
  \Delta^{21}_{\mathbf{k}} =& Tr\sum_{\mathbf{k}^{\prime}}[V^{2111}_{\mathbf{k-k^{\prime}}}e^{-\beta H_{MF}}f_{\mathbf{k}^{\prime},1}f_{-\mathbf{k}^{\prime},1}\nonumber\\
  &+V^{2122}_{\mathbf{k-k^{\prime}}}e^{-\beta H_{MF}}f_{\mathbf{k}^{\prime},2}f_{-\mathbf{k}^{\prime},2}]. \label{eq38nnb}
\end{eqnarray}
Moreover, we can prove from the equations of the interaction vertices that $V^{1112}_{\mathbf{k-k^{\prime}}}=V^{2212\star}_{\mathbf{k-k^{\prime}}}$ and $V^{1121}_{\mathbf{k-k^{\prime}}}=V^{2221\star}_{\mathbf{k-k^{\prime}}}$, inserting which into $\Delta^{11}_{\mathbf{k}}$ and $\Delta^{22}_{\mathbf{k}}$, we know that $\Delta^{11}_{\mathbf{k}}=\Delta^{22\star}_{\mathbf{k}}$. Similarly, from $V^{1211}_{\mathbf{k-k^{\prime}}}=-V^{2111}_{\mathbf{k-k^{\prime}}}$ and $V^{1222}_{\mathbf{k-k^{\prime}}}=-V^{2122}_{\mathbf{k-k^{\prime}}}$, it is obvious to obtain $\Delta^{12}_{\mathbf{k}}=-\Delta^{21}_{\mathbf{k}}$. Further considering the p-wave feature of the interaction vertex, it is straightforward to arrive at the mean-field ansatz,
\begin{eqnarray}
  \hat{\Delta}^{11}_{\mathbf{k}} &=& \Delta_{0x\mathbf{k}}-i\Delta_{0y\mathbf{k}}, \\
  \hat{\Delta}^{12}_{\mathbf{k}} &=& \Delta_{3\mathbf{k}},
\end{eqnarray}
with  $\Delta_{0x\mathbf{k}}=\Delta_{0k}\frac{k_x}{k}$ and $\Delta_{0y\mathbf{k}}=\Delta_{0k}\frac{k_y}{k}$.

\section{Self-consistent mean-field solution}

We have derived the mean-field Hamiltonian  of the CS superconductor in square lattice (Eq.\eqref{eq39s}), in this appendix we show the details in the self-consistent solution of the corresponding order parameters. The spectrum of the CS superconductor can be obtained from the mean-field Hamiltonian as $E=\pm E^{(a)}_{\mathbf{k}}$, with
\begin{equation}\label{eq86s}
 E^{(a)}_{\mathbf{k}}=\sqrt{\Delta^2_{3k}+|\mathbf{\Delta}_0+av_F\mathbf{k}|^2},
\end{equation}
where $a=\pm$. For the square lattice, where the Cooper pairs within $Q^{+}$ (or $Q^{-}$) points are favored, we have obtained the equation for the pairing potential through mean-field decoupling, i.e.,
\begin{equation}\label{eq87s}
  \Delta^{\alpha\alpha^{\prime}}_{\mathbf{k}}=\sum_{\mathbf{k}^{\prime}}V^{\alpha\alpha^{\prime}\beta^{\prime}\beta}_{\mathbf{k-k^{\prime}}}\langle f^{(+)}_{\mathbf{k}^{\prime}\beta}f^{(+)}_{-\mathbf{k}^{\prime}\beta^{\prime}}\rangle.
\end{equation}
where $V^{\alpha\alpha^{\prime}\beta^{\prime}\beta}_{\mathbf{q}}$ satisfies Eq.\eqref{eq38ma}. Then, we define the total energy of the filled band as $W=\sum_{\mathbf{k},a=\pm} E^{(a)}_{\mathbf{k}}$, which is related to the order parameter through a variation $\langle f^{(+)}_{\mathbf{k}^{\prime}\beta}f^{(+)}_{-\mathbf{k}^{\prime}\beta^{\prime}}\rangle=\frac{\delta{W}}{\delta\Delta^{\beta\beta^{\prime}}_{\mathbf{k}}}$.
Using the variation equation, the self-consistent equation is formally obtained as,
\begin{equation}\label{eq89s}
  \Delta^{\alpha\alpha^{\prime}}=\frac{1}{2}\sum_{\beta\beta^{\prime}\mathbf{k}^{\prime}}V^{\alpha\alpha^{\prime}\beta^{\prime}\beta}_{\mathbf{k-k^{\prime}}}\frac{\delta{W}}{\delta\Delta^{\beta\beta^{\prime}}_{\mathbf{k}}}.
\end{equation}
We remind that for the square lattice we have $\Delta^{12}_{\mathbf{k}}=\Delta_{3k}$ and $\Delta^{11}_{\mathbf{k}}=\Delta_{0k}\frac{k^-}{k}$.
Hence two order parameters $\Delta_{3k}$ and $\Delta_{0k}$ need to be self-consistently determined. Inserting the functional $W$ into the self-consistent eqaution, two self-consistent equations for $\Delta_{0k}$ and $\Delta_{3k}$ can be derived as following,
\begin{widetext}
\begin{eqnarray}
  \Delta_{3k} &=& \frac{\pi ev_F}{4}\sum_{\mathbf{k}^{\prime},a=\pm}\frac{A^{+}_{\mathbf{k-k^{\prime}}}k^{\prime-}+A^{-}_{\mathbf{k-k^{\prime}}}k^{\prime+}}{k^{\prime}}\frac{\Delta_{0k^{\prime}}+av_Fk^{\prime}}{E^{(a)}_{k^{\prime}}}, \\ \label{eq89na}
  \Delta_{0k}\frac{k^+}{k} &=& \frac{\pi ev_F}{2}\sum_{\mathbf{k}^{\prime},a=\pm} A^{-}_{\mathbf{k-k^{\prime}}}\frac{\Delta_{3k^{\prime}}}{E^{(a)}_{k^{\prime}}},  \label{eq89nb}
\end{eqnarray}
\end{widetext}
where $k^{\pm}$ is defined as $k^{\pm}=k_x\pm ik_y$. The sum of lattice momentum $\mathbf{k}$ can be converted into integration. After writing the integration in the sphere coordinate and then performing integration over the polar angle $\phi$, one further obtains the following simplified equations,

\begin{eqnarray} \label{eq89nna2}
  \Delta_{0k} &=& \frac{ev_F}{4}\sum_{a=\pm}\int^k_0dk^{\prime}k^{\prime}\frac{\Delta_{3k^{\prime}}}{kE^a_{k^{\prime}}}, \\   \label{eq89nnaa}
  \Delta_{3k} &=& \frac{ev_F}{4}\sum_{a=\pm}\int^{\Lambda}_kdk^{\prime}\frac{av_Fk^{\prime}+\Delta_{0k^{\prime}}}{E^a_{k^{\prime}}}.  \label{eq89nnb}
\end{eqnarray}
Solving the above equation, we can obtain the asymptotic solution for $\Delta_{0k}$ at small $k$, $\Delta_{0k}=ev_Fk/4$. To the first order in $k$ the self-consistent equation for $\Delta_{3k}$ acquires the form
\begin{equation}\label{eq90s}
\begin{split}
  \Delta_{3}&=\frac{ev_F}{4}\int^{\Lambda}_0dk^{\prime}[\frac{(e/4+1)v_Fk^{\prime}}{\sqrt{(e/4+1)^2v^{\prime2}_Fk^{\prime2}+\Delta^2_3}}\\
  &+\frac{(e/4-1)v_Fk^{\prime}}{\sqrt{(e/4-1)^2v^{\prime2}_Fk^{\prime2}+\Delta^2_3}}]
\end{split}
\end{equation}
After integration over $k^{\prime}$ and introducing dimensionless variable $u=\Lambda v_F/\Delta_3$, the solution of the above equation can be given by the crossing point of the function $F(e,u)$ with $y=0$, where $F(e,u)$ is defined as,
\begin{equation}\label{eq91s}
\begin{split}
  F(e,u)&=\frac{e}{4}[\frac{1}{e/4+1}(\sqrt{(e/4+1)^2u^2+1}-1)\\
  &+\frac{1}{e/4-1}(\sqrt{(e/4-1)^2u^2+1}-1)]-1,
\end{split}
\end{equation}
from which one can obtain the critical value of the CS charge $e_c=4/\sqrt{3}$. In the linear approximation, for $e>e_c$, we have nontrivial solution of order parameters.  In general, we can also solve the self-consistent equations by expanding the order parameter in terms of $k$. To the first order of $k$, we obtain the following general form of order parameters,
\begin{eqnarray}
  \Delta_{0k} &=& yv_Fk, \\  \label{eq91na}
  \Delta_3 &=& x\Lambda v_F,   \label{eq91nb}
\end{eqnarray}
While to second order of $k$, the general formal reads as,
\begin{eqnarray} \label{eq92na}
  \Delta_{0k} &=& y^{(0)}v_Fk, \\   \label{eq92naa}
  \Delta_3 &=& x^{(0)}(1-x^{(2)}\frac{k^2}{\Lambda^2})\Lambda v_F,  \label{eq92nb}
\end{eqnarray}
where $\Lambda$ is the momentum cutoff. For any given CS charge $e$, it is then straightforward to insert above order parameters into the self-consistent equation and  obtain the coefficients $x^{(0)}$, $x^{(2)}$, and $y^{(0)}$. These parameters are dependent on $e$. Up to the second order in $k$, $\Delta_{0k}$ can be solved giving $\Delta_{0k}=\frac{ev_Fk}{4}$. Then Inserting Eq.\eqref{eq92nb} into Eq.\eqref{eq90s}, calculating the integral on the right hand side of the equation and expanding the expression up to $k^2$, we obtain  $x^{(0)}=e/4$ and $x^{(2)}=1$ for the square lattice.


\section{Details of the derivation of the Nambu-Goldstone mode}

\subsection{Square lattice}

We associate a $U(1)$ phase to the matrix order parameter and cast the Hamiltonian in the form
\begin{equation}\label{eq73s}
\begin{split}
  H_{MF}&=v_F\boldsymbol{\sigma}\cdot\mathbf{k}\frac{\tau^3+1}{2}+v_F\boldsymbol{\sigma}^{T}\cdot\mathbf{k}\frac{1-\tau^3}{2}+\hat{\Delta}_{\mathbf{k}}e^{2i\theta}\tau^+\\
  &+\hat{\Delta}^{\dagger}_{\mathbf{k}}e^{-2i\theta}\tau^-.
\end{split}
\end{equation}
Then, the imaginary-time action in functional integral representation can be written as,
\begin{equation}\label{eq74s}
\begin{split}
  S&=\int d\tau\sum_{\mathbf{k}}\psi^{\dagger}_{\mathbf{k}}[\partial_{\tau}-v_F\boldsymbol{\sigma}\cdot\mathbf{k}\frac{\tau^3+1}{2}-v_F\boldsymbol{\sigma}^{T}\cdot\mathbf{k}\frac{1-\tau^3}{2}\\
  &-\hat{\Delta}_{\mathbf{k}}e^{2i\theta}\tau^+-\hat{\Delta}^{\dagger}_{\mathbf{k}}e^{-2i\theta}\tau^-]\psi_{\mathbf{k}}.
\end{split}
\end{equation}
After performing a gauge transformation, $\psi_{\mathbf{k}}\rightarrow\psi^{\prime}_{\mathbf{k}}=U\psi_{\mathbf{k}}$, with $U=e^{-i\theta}(\tau^3+1)/2+e^{i\theta}(1-\tau^3)/2$, the inverse of the Green's function of the Chern-Simons superconductor acquires the following matrix form,
\begin{equation}\label{eq75s}
  G^{-1}_{\mathbf{k},\theta}=\left(
                               \begin{array}{cc}
                                 e^{-i\theta}(\partial_{\tau}-v_F\boldsymbol{\sigma}\cdot\mathbf{k})e^{i\theta} & -\hat{\Delta}_{\mathbf{k}} \\
                                 -\hat{\Delta}_{\mathbf{k}}^{\dagger} & e^{i\theta}(\partial_{\tau}-v_F\boldsymbol{\sigma}^T\cdot\mathbf{k})e^{-i\theta} \\
                               \end{array}
                             \right),
\end{equation}
which can be further simplified after taking into account the noncommunity between $\mathbf{k}$ and the $\theta$ field, which is in general dependent on spatial coordinates.

We further decompose the Green's function into two parts, according to whether or not it contains the phase field $\theta$,
\begin{equation}\label{eq77s}
  G^{-1}_{\mathbf{k},\theta}=[G^{0}_{\mathbf{k}}]^{-1}-\Sigma_{\theta},
\end{equation}
The $\theta$-independent free part reads as,
\begin{equation}\label{eq78s}
  [G^{0}_{\mathbf{k}}]^{-1}=\left(
                               \begin{array}{cc}
                                 \partial_{\tau}-v_F\boldsymbol{\sigma}\cdot(-i\boldsymbol{\nabla}) & -\hat{\Delta}_{\mathbf{k}} \\
                                 -\hat{\Delta}_{\mathbf{k}}^{\dagger} & \partial_{\tau}-v_F\boldsymbol{\sigma}^T\cdot(-i\boldsymbol{\nabla}) \\
                               \end{array}
                             \right).
\end{equation}
while the phase-dependent self-energy enjoys the form,
\begin{equation}\label{eq79s}
  \Sigma_{\theta}=\left(
                               \begin{array}{cc}
                                 -i\partial_{\tau}\theta+v_F\boldsymbol{\sigma}\cdot\nabla\theta & 0 \\
                                 0 & i\partial_{\tau}\theta-v_F\boldsymbol{\sigma}^T\cdot\nabla\theta \\
                               \end{array}
                             \right).
\end{equation}
Then one can expand the action $S_{\theta}$ in terms of $\Sigma_{\theta}$. We are interested in the leading order that has nonvanishing contributions. Up to the second order in the derivatives, the action reads as,
\begin{equation}\label{eq80s}
\begin{split}
  S_{\theta}&=-\mathrm{Trln}G^{-1}_{\mathbf{k}}=-\mathrm{Trln}[G^{0}_{\mathbf{k}}]^{-1}+\mathrm{Tr}(G^{0}_{\mathbf{k}}\Sigma_{\theta})\\
  &+\frac{1}{2}\mathrm{Tr}(G^{0}_{\mathbf{k}}\Sigma_{\theta}G^{0}_{\mathbf{k}}\Sigma_{\theta}).
\end{split}
\end{equation}
Obviously, the first term is a constant of the trace, which is irrelevant to the phase fluctuation in Chern-Simons superconductivity. As this stage we calculate the second and the third-order terms in Eq.\eqref{eq80s}. We first take the inverse of Eq.\eqref{eq78s}, which leads to
\begin{widetext}
\begin{equation}\label{eq81s}
\begin{split}
  G^{0}_{\mathbf{k}}&=\frac{1}{A}[a\sigma^3\tau^3-b\sigma^0\tau^0-(c\sigma^++c^{\star}\sigma^-)(\frac{\tau^3+1}{2})
  -(c^{\star}\sigma^++c\sigma^-)(\frac{-\tau^3+1}{2})+ie\sigma^1\tau^2+d\sigma^2\tau^2]\\
  &-(f(\frac{\sigma^3+1}{2})+f^{\star}(\frac{1-\sigma^3}{2}))\tau^+-(f^{\star}(\frac{\sigma^3+1}{2})
  +f(\frac{1-\sigma^3}{2}))\tau^-],
\end{split}
\end{equation}
\end{widetext}
where $A$ is the determinant of $[G^{0}_{\mathbf{k}}]^{-1}$ defined by,
\begin{equation}\label{eq82s}
  A=(\Delta_{3k}^2+|\boldsymbol{\Delta}_0-v_F\mathbf{k}|^2+\omega^2)(\Delta_{3k}^2+|\boldsymbol{\Delta}_0+v_F\mathbf{k}|^2+\omega^2),
\end{equation}
 The variables $a,b,c,d,e,f$ are defined as,
\begin{widetext}
\begin{equation} \label{eq83sss}
  \begin{split}
  a &= 2\Delta_{3k}v_F\mathbf{\Delta}_0\cdot\mathbf{k},    \\
  b &= i\omega(\Delta^2_{3k}+|\mathbf{\Delta}_{0}|^2+k^2v^{2}_F+\omega^2),     \\
  c &= v_F(\Delta_{0yk}+i\Delta_{0xk})^2k^++v_F(\Delta^2_{3k}+\omega^2+v^{2}_Fk^2)k^-,  \\
  d &= \Delta^3_{3k}+\Delta_{3k}(|\mathbf{\Delta}_0|^2+k^2v^{2}_F+\omega^2),   \\
  e &= 2i\mathbf{\Delta}_0\cdot\mathbf{k}v_F\omega,   \\
  f &=
  (|\mathbf{\Delta}_0|^2+\Delta^2_{3k}+\omega^2)(\Delta_{0xk}-i\Delta_{0yk})-v^{2}_F(\Delta_{0xk}+i\Delta_{0yk})(k^-)^2,
  \end{split}
\end{equation}
\end{widetext}
Rewriting $\Sigma_{\theta}$ in terms of the Pauli matrices, we obtain
\begin{equation}\label{eq83s}
  \Sigma_{\theta}=(-i\partial_{\tau}\theta)\tau^3+v_F\boldsymbol{\sigma}\cdot\nabla\theta(\frac{1+\tau^3}{2})-v_F\boldsymbol{\sigma}^T\cdot\nabla\theta(\frac{1-\tau^3}{2}).
\end{equation}
Then $\mathrm{Tr}G^{0}_{\mathbf{k}}\Sigma_{\theta}$ can be calculated analytically. After expansion, it can be clearly seen that the first order term becomes zero after the trace over the Pauli matrices. Therefore, we only need to consider the second-order contribution.
Furthermore, it is straightforward to prove the following identities for integrals:
\begin{equation}\label{eq84na}
\begin{split}
  \int^{2\pi}_0 d\phi c^2 &= \int^{2\pi}_0 d\phi c^{\star2}=0, \\  
  \int^{2\pi}_0 d\phi f^2 &= \int^{2\pi}_0 d\phi f^{\star2}=0, \\  
  \int^{2\pi}_0 d\phi bc  &= \int^{2\pi}_0 d\phi bc^{\star}=0.  
\end{split}
\end{equation}
Using the above relations, most of the terms after expansion will vanish after integration over $\phi$. This is nothing but a direct manifestation of the rotational symmetry. Finally, we obtain the action of the Goldstone mode in the Chern-Simons superconductor on a square lattice in the form
\begin{equation}\label{eq85s}
  S_{\theta}=\int d\tau d\mathbf{r}[c_1(i\partial_{\tau}\theta)^2+c_2(\nabla\theta)^2],
\end{equation}
where
\begin{eqnarray} \label{eq85na}
  c_1 &=& \mathrm{tr}[\frac{2}{A^2}(a^2+b^2+|c|^2+e^2-|f|^2-d^2)], \\   \label{eq85naa}
  c_2 &=& \mathrm{tr}[\frac{2}{A^2}v^{2}_F(-a^2+b^2+d^2+e^2)].   \label{eq85nb}
\end{eqnarray}
where ``$\mathrm{tr}$" denotes $\mathrm{tr}[...]=\int^{\infty}_{-\infty} d\omega\int^{\Lambda_0}_{0} dkk\int^{2\pi}_0d\phi$. $\Lambda_{0}=\pi/2a$ should be set to include half of the Brillouin zone.

From Eq.\eqref{eq85s}, we can obtain the dispersion of phase fluctuation of the CS superconductor. Constants $c_1$ and $c_2$ are dependent on $a$, $b$, $c$, $d$, $e$, $f$, which are in turn dependent on the order parameters $x^{(0)}$, $x^{(2)}$, and $y^{(0)}$ in Eq.\eqref{eq92na}-\eqref{eq92nb}. Therefore, the dispersion is a function of the CS charge $e$. For a given $e$, $x^{(0)}$, $x^{(2)}$, and $y^{(0)}$ have been obtained up to the quadratic approximation (Eq.\eqref{eq92na}-\eqref{eq92nb}), and the order parameters have been found as
\begin{eqnarray} \label{eq85nna}
  \Delta_{0k} &=& \frac{ev_Fk}{4}, \\   \label{eq85nnaa}
  \Delta_3 &=& \frac{e}{4}(1-\frac{k^2}{\Lambda^2})\Lambda v_F.   \label{eq85nnb}
\end{eqnarray}
Inserting Eq.\eqref{eq85nna}-\eqref{eq85nnb} into Eq.\eqref{eq85s}-\eqref{eq85nb} one obtains the dispersion of the Nambu-Goldstone mode as a function of $e$.

\subsection{Honeycomb lattice}
We have also performed the above computation also in the case of the honeycomb lattice. Since almost all the steps are similar to the square lattice case, we do not present the details here. The final action for the phase fluctuations is found to have the same form as Eq.\eqref{eq85s}-\eqref{eq85nb}. However, the self-consistent equations differ from Eq.\eqref{eq89nnb}, which leads to the following solution of order parameters on the honeycomb lattice up to quadratic order in $k$,
 \begin{eqnarray}
  \Delta_{0k} &=& \frac{ev^{\prime}_Fk}{2}, \\   \label{eq85nnna}
  \Delta_3 &=& \frac{e}{2}(1-\frac{k^2}{\Lambda^2})\Lambda v^{\prime}_F£¬   \label{eq85nnnb}
\end{eqnarray}
with $v^{\prime}_F=\frac{\sqrt{3}\epsilon J}{2}$. For small CS charge $e$, where the CS superconductor has just been formed, it is found that there is some quantitative deviation between the two lattices. This corresponds to the strong-fluctuating region, where different lattice symmetries still play some role in shaping of the Nambu-Goldstone mode. However, for the physically stable phase at large $e$, the Goldstone modes in the two lattices are found to saturate  at almost the same finite value.  The results for the honeycomb lattice are plotted in Fig.2(b) of the main text. It clearly shows the lattice-independence of our theory.

\subsection{The result}

Finally, it is worthwhile to note that even though the solution up to $k^2$ expansion works very well, it still has minor deviations from the lattice results\cite{Sedrakyanc}. In order to capture this deviation, we can introduce the ratio between the momentum cutoff $\Lambda$ and $\Lambda_0$ as $\alpha=\frac{\Lambda}{\Lambda_0}=\frac{\Lambda}{(\pi/2a)}$ (for square lattice). Then, after making the integral in Eq.\eqref{eq85na} and Eq.\eqref{eq85nb} dimensionless, $c_2/c_1$ can be proved to only explicitly dependent on $\alpha$. The dispersion then is finally obtained as,
\begin{equation}\label{eq65s}
  \omega=uk,
\end{equation}
where the coefficient $u$ is
\begin{equation}\label{eq66s}
  u=\sqrt{\frac{c_2}{c_1}}=v_F\sqrt{\frac{I_2}{I_1}}.
\end{equation}
The function $I_1(\alpha)$ reads as,
\begin{widetext}
\begin{equation}\label{eq63s}
\begin{split}
  I_1(\alpha) &=\int^1_0dk^{\prime}k^{\prime}\int^{\infty}_{-\infty} d\omega^{\prime}\frac{1}{[\alpha^2x^2+(y-1)^2k^{\prime2}+\omega^{\prime2}]^2[\alpha^2x^2+(y+1)^2k^{\prime2}+\omega^{\prime2}]^2}\\
     &\times\{4y^2k^{\prime4}(\alpha^2x^2-\omega^{\prime2})-(\omega^{\prime2}+\alpha^2x^2)[\alpha^2x^2+(y^2+1)k^{\prime2}+\omega^{\prime2}]^2+k^{\prime2}[(1-y^2)k^{\prime2}+\alpha^2x^2+\omega^{\prime2}]^2\\
     &-y^2k^{\prime2}[(y^2-1)k^{\prime2}+\alpha^2x^2+\omega^{\prime2}]^2\},
     \end{split}
\end{equation}
and $I_1(\alpha)$ reads as,
\begin{equation}\label{eq64s}
\begin{split}
  I_2(\alpha) &= \int^1_0dk^{\prime}k^{\prime}\int^{\infty}_{-\infty} d\omega^{\prime}\frac{1}{[\alpha^2x^2+(y-1)^2k^{\prime2}+\omega^{\prime2}]^2[\alpha^2x^2+(y+1)^2k^{\prime2}+\omega^{\prime2}]^2}\\
     &\times\{-4y^2k^{\prime4}(\alpha^2x^2+\omega^{\prime2})+(\alpha^2x^2-\omega^{\prime2})[\alpha^2x^2+(y^2+1)k^{\prime2}+\omega^{\prime2}]^2\}.
\end{split}
\end{equation}
\end{widetext}
The saturated value of the linearity coefficient is found to be very weakly dependent on $\alpha$.  The undetermined higher order contribution $O(k^3)$ in the self-consistent equation can be taken into account by allowing deviation of $\alpha$ from $1$. In this way, one can evaluate the error bar in the dispersion $u$.  For a safe lower and upper bound of $\Lambda$ with $\alpha\in[0.3,1]$, the estimated error bar gives the variation region of the saturated linearity coefficient.  The region is obtained as $[0.487,0.535]$ for the square lattice (in unit of $v_F$), and $[0.487,0.532]$ for the honeycomb lattice (in unit of $v^{\prime}_F$). This is in high-precision agreement with the dispersion of magnons with coefficient $0.5$.

\section{Anyonic statistics via CS fermion representation}
We consider the braiding between two states, $\bar{S}^{\dagger}_i|0\rangle$ and $\bar{S}^{\dagger}_j|0\rangle$. In second quantized language, we first consider the hopping of a particle from $\mathbf{r}$ to $\mathbf{r}^{\prime}$, i.e., $\bar{S}^{\dagger}_{\mathbf{r}^{\prime}}\bar{S}_{\mathbf{r}}$. Inserting the CS fermion representation,
$\bar{S}^{\pm}_{\mathbf{r}}=f^{\pm}_{\mathbf{r}}U^{\pm}_{\mathbf{r}}$,
we have $\bar{S}^{\dagger}_{\mathbf{r}^{\prime}}\bar{S}_{\mathbf{r}}=f^{\dagger}_{\mathbf{r}^{\prime}}e^{i\phi_{\mathbf{r}^{\prime},\mathbf{r}}}f_{\mathbf{r}}$, where $\phi_{\mathbf{r}^{\prime},\mathbf{r}}=e\sum_{\mathbf{r}^{\prime\prime}}[\mathrm{arg}(\mathbf{r}
-\mathbf{r}^{\prime\prime})-\mathrm{arg}(\mathbf{r}^{\prime}-\mathbf{r}^{\prime\prime})]n_{\mathbf{r}^{\prime\prime}}
=\chi_{\mathbf{r}}-\chi_{\mathbf{r}^{\prime}}$ and $\chi_{\mathbf{r}}=e\sum_{\mathbf{r}^{\prime\prime}}\mathrm{arg}(\mathbf{r}
-\mathbf{r}^{\prime\prime})n_{\mathbf{r}^{\prime\prime}}$.
Obviously, $\phi_{\mathbf{r}^{\prime},\mathbf{r}}$ is the CS flux attached to the fermions. The process where a  $\bar{S}$-particle  hops from $\mathbf{r}$ to $\mathbf{r}^{\prime}$ is equivalent to an $f$-fermion that moves from $\mathbf{r}$ to $\mathbf{r}^{\prime}$ and with an additional phase of $\phi_{\mathbf{r}^{\prime},\mathbf{r}}$. To study the statistical angle $\gamma$ and the braiding between the two states $\bar{S}^{\dagger}_i|0\rangle$ and $\bar{S}^{\dagger}_j|0\rangle$, we can calculate, in the equivalent picture, the total phase accumulated by the $f$-fermions and add their own statistical angle of $\pi$.

To this end, let us consider the $f^{\dagger}_i|0\rangle$ state which adiabatically circles around another state $f^{\dagger}_j|0\rangle$ forming a closed loop in real space. The total contribution of the Berry phase from the CS attachment is then the phase $\phi_{\mathbf{r}^{\prime},\mathbf{r}}$ with $\mathbf{r}=\mathbf{R}_i$ and $\mathbf{r}^{\prime}=\mathbf{R}_i$, i.e.,

\begin{equation}\label{eqss65}
  \phi_{i,i}=\oint d\mathbf{r}\cdot\nabla_{\mathbf{r}}\chi_{\mathbf{r}}=
  e\sum_{\mathbf{r}^{\prime\prime}}n_{\mathbf{r}^{\prime\prime}}\oint d\mathbf{r}\cdot\nabla\mathrm{arg}(\mathbf{r}-\mathbf{r}^{\prime\prime}).
\end{equation}
Moreover, since $\nabla\mathrm{arg}(\mathbf{r}-\mathbf{r}^{\prime\prime})=\hat{z}\times
(\mathbf{r}-\mathbf{r}^{\prime\prime})/|\mathbf{r}-\mathbf{r}^{\prime\prime}|^2$, $\phi_{i,i}$ can be calculated to be $\phi_{i,i}=2\pi e\sum_{\mathbf{r}^{\prime\prime}}n_{\mathbf{r}^{\prime\prime}}$. Here $\sum_{\mathbf{r}^{\prime\prime}}n_{\mathbf{r}^{\prime\prime}}$ is the total number  fermionic $f$-states enclosed by the closed trajectory. As we consider the braiding between two states, the loop will only enclose a single state, giving $\sum_{\mathbf{r}^{\prime\prime}}n_{\mathbf{r}^{\prime\prime}}=1$. The braiding between two states is equivalent to the process where the $f^{\dagger}_i|0\rangle$ state completes half of the trajectory loop enclosing the $f^{\dagger}_j|0\rangle$ state, therefore the contribution during the braiding from the CS attachment is half of $\phi_{i,i}$, i.e., $e\pi$.
Besides the CS attachment, the braiding between the two fermionic states, itself, gives rise to a $\pi$ phase. Thus, the total statistical angle amounts to $\gamma=\pi+e\pi$. It is clear that  the CS fermion representation generates hardcore-bosonic models only when $e=2l+1$, while for a generic $e$ or for $0<e<1$, operators $\bar{S}^+$ and $\bar{S}^-$ represent anyonic operators with tunable statistical angle. In this case, the model $H=J\sum_{\langle ij\rangle}\bar{S}^x_i\bar{S}^x_j+\bar{S}^y_i\bar{S}^y_j$ describes the anyonic gas.



\begin{thebibliography}{99}
\bibitem{footnote} In Ref. \onlinecite{Sedrakyanc} the existence of a spin-nematic-type long-range order in the CS superconductor is shown,
but the existence of N\'{e}el order requires calculation of two-point correlation functions, which is out of the scope of the present project.
\end{thebibliography}
\end{document}